\documentclass[11pt, a4paper]{article}
\usepackage[dvipdfmx]{graphicx}
\usepackage{setspace,amsmath,amssymb,amsthm}
\usepackage{bm}
\usepackage{bbm}
\usepackage{enumitem,color}
\usepackage{mathrsfs}
\usepackage{physics}
\usepackage{fancybox}
\usepackage{algorithm}
\usepackage{algorithmic}
\usepackage{natbib}
\usepackage{url}
\usepackage{booktabs}
\usepackage{comment}
\usepackage{caption}
\usepackage{setspace}

\usepackage[top=25mm,bottom=27mm,left=25mm,right=25mm]{geometry}

\newtheorem*{rem*}{Remark}

\newcommand{\true}[1]{{#1}^{\text{true}}}
\newcommand{\E}[2]{\mathbb{E}_{#1}[#2]}

\numberwithin{equation}{section}

\title{\vspace{-25mm}Bayesian Stochastic Block Models for Weighted Network Data Based on the Zero-Inflated Negative Binomial Distribution}
\author{Fumiya Iwashige \\
Department of Mathematics, Hiroshima University}
\date{April 22, 2026}

\begin{document}
\maketitle
\doublespacing
\begin{abstract} 
Weighted networks encode not only the presence of interactions but also their strength. Existing methods for weighted network community detection often rely on Poisson models. However, this can be restrictive for overdispersed data and make efficient posterior computation difficult when covariates are incorporated. We propose Bayesian stochastic block models based on the zero-inflated negative binomial distribution: ZINB-SBM without covariates and CZINB-SBM with pairwise covariates. The proposed models accommodate overdispersion, naturally account for missing interactions through zero inflation, and admit efficient Gibbs sampling. In CZINB-SBM, P\'{o}lya-Gamma data augmentation enables posterior inference for regression coefficients with uncertainty quantification. We further employ a dynamic mixture of finite mixtures, which allows the number of communities to be inferred from the data and can lead to more accurate clustering. Simulation studies show that ZINB-SBM is more robust than a zero-inflated Poisson SBM for highly overdispersed networks. Real data analysis demonstrates interpretable block specific covariate effects and substantially improved missing link prediction compared with a Poisson regression-based Bayesian SBM.

\end{abstract}


\section{Introduction}\label{sec:intro}
Network data have been widely studied in recent years across a diverse range of fields, including sociology, ecology, and biology. In particular, much attention has been devoted to community detection, a clustering problem for network data that aims to classify nodes into subgroups with dense connections within groups and sparse connections between groups. In a model-based approach, the network generating process is described using a probabilistic model, which provides an intuitive and tractable way to interpret and infer the network structure. Furthermore,  the use of Bayesian statistics facilitates uncertainty quantification for parameters of interest through the posterior distribution. 

Many networks are observed as weighted networks, in which edges encode the strength or frequency of interactions between nodes. Accordingly, it is desirable to model such data without discarding the information in the weights. While Poisson distributions are widely used for modeling weighted network data \citep{Luetal10.1093/jrsssa/qnaf029,mariadassou2010uncovering}, they impose the restrictive assumption that the mean equals the variance, making them unsuitable for overdispersed network data. Moreover, in real-world network data, interactions that should have been observed may be missing because of external factors or measurement errors. These missing edges can substantially affect the results of the analysis. In such situations, the missing edges may be predicted using external information such as covariates. However, Bayesian methods based on Poisson regression make it difficult to construct efficient MCMC algorithms \citep{mariadassou2010uncovering}. In this study, we adopt a Bayesian approach to the stochastic block model (SBM), a standard statistical model for model-based community detection \citep{Nowicki01092001}, and propose ZINB-SBM, an SBM based on the zero-inflated negative binomial (ZINB) distribution \citep{neelon2019bayesian}. By using the negative binomial distribution, the proposed model can accommodate overdispersion and, even when covariates are incorporated into the likelihood function, allows the construction of a Gibbs sampler through P\'{o}lya-Gamma data augmentation. The zero-inflation structure also makes it possible to incorporate missing data into the model. 

Two novel aspects of our model are the use of a dynamic mixture of finite mixtures (DMFM) \citep{Fruhwirthetal/21-BA1294,iwashige2025bayesianmixturemodelingusing} and the incorporation of pairwise covariates into the likelihood function. Since a prior distribution is assigned to the number of communities, DMFMs allow the number of communities to be inferred from the data. However, this feature is not unique to DMFMs, as the number can likewise be inferred in MFMs \citep{Miller02012018,argiento2022infinity} and Dirichlet process mixtures \citep{Escobar1995}. As discussed in \cite{Fruhwirthetal/21-BA1294,Greveetal2022}, the resulting prior on partitions induced by DMFMs can be more flexible than those induced by conventional MFMs and Dirichlet process mixtures, which may lead to improved clustering accuracy. By contrast, \cite{Lu2023BayesianStrategies} proposed an SBM based on the zero-inflated negative binomial distribution, where the number of communities was selected using an information criterion. In addition, our model incorporates pairwise covariates into the likelihood via logistic functions. The approach of \cite{mariadassou2010uncovering} is based on Poisson regression. Although \cite{Luetal10.1093/jrsssa/qnaf029} showed that node attributes can be used as external information to improve label prediction through a prior on the number of communities, incorporating covariates directly into the likelihood is more desirable when the goal is to predict interactions. By contrast, covariates are not incorporated in \cite{Lu2023BayesianStrategies}.

This paper is organized as follows. In Section~\ref{sec:metho}, we present the proposed models ZINB-SBM and CZINB-SBM. In Section~\ref{sec:PostComp}, we give an overview of posterior computation for the proposed models, but the details of which are provided in the Appendix. In Sections~\ref{sec:sim} and~\ref{sec:real}, we demonstrate the usefulness of the proposed models through synthetic and real data analyses. \texttt{R} code implementing the proposed method is available at the GitHub repository \url{https://github.com/Fumiya-Iwashige/ZINB-SBM_CZINB-SBM}.

\section{Methods}\label{sec:metho}
\subsection{ZINB-SBM without covariates}
Let \(A\) be an \(n \times n\) adjacency matrix of a weighted network with \(n\) nodes. Each element \(A_{ij}\) is a nonnegative integer representing the edge between nodes \(i\) and \(j\). In particular, \(A_{ij}=0\) indicates that there is no relationship between nodes \(i\) and \(j\), whereas \(A_{ij}>0\) indicates not only the presence of a relationship but also the strength of the interaction between them. We further assume that the network is undirected and has no self-loops, that is, \(A_{ij}=A_{ji}\) and \(A_{ii}=0\). Our proposed ZINB-SBM is defined by the following Bayesian hierarchical model:
\begin{align}\label{model:ZINB}
    \begin{split}
     &K-1 \sim q_{K},\quad q_{K}\ \text{is a probability mass function on}\ \mathbb{Z}_{\geq 0},\\
     &z_i \mid \pi, K \overset{\mathrm{i.i.d.}}{\sim} \mathrm{Categorical}(\pi),\quad i =1,\dots,n,\\
    &\pi \mid \gamma, K \sim \mathrm{Dirichlet}_K(\gamma/K,\dots,\gamma/K),\\
    &A_{ij} \mid z_i = l, z_j = m, p_{lm}, \psi_{lm}, r_{lm} \overset{\mathrm{ind}}{\sim} \mathrm{ZINB}(p_{lm}, \psi_{lm}, r_{lm}),\quad 1 \leq i < j \leq n,\\
    &p_{lm} \overset{\mathrm{i.i.d.}}{\sim} \mathrm{Beta}(a_p, b_p),\quad \psi_{lm} \overset{\mathrm{i.i.d.}}{\sim} \mathrm{Beta}(a_{\psi}, b_{\psi}),\quad r_{lm} \overset{\mathrm{i.i.d.}}{\sim} \mathrm{Gamma}(a_r, b_r),\quad 1 \leq l \leq m \leq K,
    \end{split}
\end{align}
where \(K\) is the number of communities and \(z_i\) is the community label of node \(i\). Here, \(\mathrm{ZINB}(p_{lm},\psi_{lm},r_{lm})\) denotes the zero-inflated negative binomial distribution with the following probability mass function:
\[
f(A_{ij}) = p_{lm}\mathbbm{1}\{A_{ij}=0\} + (1-p_{lm})\frac{\Gamma(A_{ij}+r_{lm})}{\Gamma(r_{lm})A_{ij}!}\psi_{lm}^{r_{lm}}(1-\psi_{lm})^{A_{ij}},
\]
where \(\mathbbm{1}\{\cdot\}\) is an indicator function, \(p\) denotes the zero-inflation probability and \(\psi\) and \(r\) are the parameters of the negative binomial component. In model \eqref{model:ZINB}, the ZINB can be represented equivalently using latent variables as follows \citep{GHOSH20061360}:
\begin{align}\label{eq:ZINB_miss}
    \begin{split}
    &A_{ij} = w_{ij}(1 - x_{ij}),\\
    &w_{ij} \mid z_i = l, z_j = m, \psi_{lm} \sim \text{NB}(\psi_{lm}, r_{lm}),\\
    &x_{ij} \mid z_i = l, z_j = m, p_{lm} \sim \text{Bernoulli}(p_{lm}).
    \end{split}
\end{align}
Here, \(w_{ij}\) represents the latent interaction between nodes \(i\) and \(j\). In addition, \(x_{ij}\) is a latent binary variable that determines whether \(A_{ij}\) is forced to be zero as a structural zero with probability \(p_{lm}\). By introducing this zero inflation structure together with latent variables, the model can account for latent interactions and missing observations.

This model has two important features for community detection. First, it is a SBM because the distribution of \(A_{ij}\) depends on nodes \(i\) and \(j\) through their latent community labels \(z_i\) and \(z_j\). Therefore, all node pairs in the same community pair \((l,m)\) share the same distributional parameters. Second, the model is based on a DMFM. Following \citet{Fruhwirthetal/21-BA1294}, we assume that \(q_K\) follows a beta-negative-binomial distribution, \(\mathrm{BNB}(\alpha_{\lambda}, a_{\pi}, b_{\pi})\), and assign an \(F\)-distribution prior \(\mathcal{F}(6,3)\) to \(\gamma\).

\subsection{Incorporating pairwise covariates}
In real network data, pairwise covariates for each node pair \((i,j)\) often influence the interaction \(A_{ij}\) \citep{Vacher_etal2008,mariadassou2010uncovering}. In such settings, it is necessary to incorporate into the model not only heterogeneity between communities but also heterogeneity between node pairs that cannot be captured by standard SBMs. Let \(y_{ij} \in \mathbb{R}^{q}\) denote a \(q\)-dimensional pairwise covariate vector for the node pair \((i,j)\). The mixture weight \(p_{ij}\) and the success probability \(\psi_{ij}\) of the zero-inflated negative binomial distribution are then modeled by logistic functions:
\begin{align}\label{eq:logistic}
    \psi_{ij} = \frac{\exp(y_{ij}^\top\beta_{z_iz_j,1})}{1 + \exp(y_{ij}^\top\beta_{z_iz_j,1})},\quad p_{ij}=\frac{\exp(y_{ij}^\top\beta_{z_iz_j,2})}{1 + \exp(y_{ij}^\top\beta_{z_iz_j,2})},\quad 1 \leq i < j \leq n,
\end{align}
where \(\beta_{lm,1}, \beta_{lm,2} \in \mathbb{R}^{q}\) are regression coefficient vectors. The subscripts \(1\) and \(2\) distinguish the regression coefficients for \(\psi\) and \(p\), respectively. Then, our proposed CZINB-SBM is defined by the following Bayesian hierarchical model (the components identical to those in \eqref{model:ZINB} are omitted):
\begin{align}\label{model:CZINB}
    \begin{split}
    &A_{ij} \mid z_i = l, z_j = m, \beta_{lm,1}, \beta_{lm,2}, r \overset{\mathrm{ind}}{\sim} \mathrm{ZINB}(p_{ij}, \psi_{ij}, r),\quad 1 \leq i < j \leq n,\\
    &\beta_{lm,s} \overset{\mathrm{i.i.d.}}{\sim} \mathcal{N}_q(b_0, B_0),\quad  1 \leq l \leq m \leq K,\quad s=1,2,\\
    &r \sim \text{Gamma}(a_r, b_r),
    \end{split}
\end{align}
where $p_{ij}$ and $q_{ij}$ are given by \eqref{eq:logistic}, and $\mathcal{N}_q(b_0, B_0)$ is \(q\)-dimensional normal distribution with mean vector $b_0$ and variance-covariance matrix $B_0$. In \eqref{model:CZINB}, the ZINB distribution can also be represented using latent variables in the same way as in \eqref{eq:ZINB_miss}; the only difference is that \(\psi_{lm}\), \(p_{lm}\), and \(r_{lm}\) are replaced with \(\psi_{ij}\), \(p_{ij}\), and \(r\), respectively.

The essential feature of CZINB-SBM lies in \eqref{eq:logistic}. Specifically, the model retains the SBM structure because the regression coefficients depend on the latent community labels of nodes \(i\) and \(j\). At the same time, by incorporating the pairwise covariates, CZINB-SBM is able to capture heterogeneity between node pairs that cannot be represented by standard SBMs. Even when interactions are potentially unobserved, pairwise covariates can be used as external information to help predict both the presence of an interaction (missing link prediction) and the weight of the underlying latent interaction.

For Bayesian computation, it follows from \eqref{eq:logistic} and \eqref{model:CZINB} that, by using P\'{o}lya-Gamma data augmentation \citep{polson2013bayesian, neelon2019bayesian}, the full conditional distributions of the regression coefficients are Gaussian. By contrast, for the method in \cite{mariadassou2010uncovering} based on Poisson regression, sampling the regression coefficients is not straightforward, so a variational algorithm is employed. For CZINB-SBM, a Gibbs sampler can be constructed, which yields posterior samples of the regression coefficients. Therefore, the proposed method enables not only point estimation but also posterior inference on the regression coefficients, including uncertainty quantification.
\begin{rem*}
   In \eqref{model:CZINB}, we do not impose a block structure on the overdispersion parameter. This is because when updating the label variables, the block structure requires evaluation of the P\'{o}lya-Gamma density function. By instead treating \(r\) as a global parameter, this evaluation is no longer needed, while the model still retains sufficient flexibility.
\end{rem*}

\section{Posterior computation}\label{sec:PostComp}
Here, we outline the Bayesian computational methods for ZINB-SBM and CZINB-SBM. We employ the extended telescoping sampler \citep{iwashige2025bayesianmixturemodelingusing} for posterior computation of DMFMs. Since ZINB-SBM and CZINB-SBM are essentially finite mixture models with a ZINB kernel, posterior computation can be divided into two parts: updating the parameters associated with the finite mixture model and updating those associated with the kernel. 

For the latter set of parameters in ZINB-SBM, \(p_{lm}\) and \(\psi_{lm}\) are sampled from Beta distributions by conjugacy. The overdispersion parameter \(r_{lm}\) is updated using a random-walk Metropolis-Hastings step. The label variable \(z_i\) is updated after marginalizing out \(p_{lm}\) and \(\psi_{lm}\). For CZINB-SBM, the block parameters to be updated are the regression coefficients. By introducing latent variables following P\'{o}lya-Gamma distributions, the full conditional distributions of the regression coefficients become Gaussian. In updating the label variables, the regression coefficients are marginalized out, while the conditional dependence on the P\'{o}lya-Gamma latent variables is retained.

\section{Simulation study}\label{sec:sim}
In this section, we compare our proposed ZINB-SBM with ZIP-SBM, a SBM based on the zero-inflated Poisson distribution \citep{Luetal10.1093/jrsssa/qnaf029}, using synthetic data from the perspective of community detection. The aim of this simulation study is to investigate, from the perspective of community detection, whether ZINB-SBM is more robust than ZIP-SBM to overdispersion in the data.

\subsection{Data generating process}
Let \(K^{\mathrm{true}} = 3\) and \(n = 150\). The true community labels \(z^{\mathrm{true}}\) are generated from a categorical distribution with parameter \(\pi^{\mathrm{true}}\), where \(\pi^{\mathrm{true}}=(1/3,1/3,1/3)\). We consider the following two ways of generating the interaction \(A_{ij}\):
\begin{align}
    &\text{Scenario}~1\quad A_{ij} \mid z_i = l,z_j = m, p_{lm}^{\ast}, \psi_{lm}^{\ast}, r_{lm}^{\ast} \overset{ind}{\sim} \text{ZINB}(p_{lm}^{\ast}, \psi_{lm}^{\ast}, r_{lm}^{\ast}),\quad 1 \leq i < j \leq n, \notag\\
    &\text{Scenario}~2\quad A_{ij} \mid z_i = l,z_j = m, p_{lm}^{\ast},\lambda_{lm}^{\ast}, \overset{ind}{\sim} \text{ZIP}(p_{lm}^{\ast}, \lambda_{lm}^{\ast}),\quad 1 \leq i < j \leq n,\notag
\end{align}
where the true block parameters \(p_{lm}^{\ast}, \psi_{lm}^{\ast}, r_{lm}^{\ast},\lambda_{lm}^\ast\) are set as follows:
\begin{align*}
    \begin{split}
    &p_{lm}^\ast = \mathbbm{1}\{l=m\}0.1 + \mathbbm{1}\{l\neq m\}0.7,\quad \psi_{lm}^\ast = \mathbbm{1}\{l=m\}0.1 + \mathbbm{1}\{l\neq m\}0.2,\\
    &r_{lm}^\ast = \mathbbm{1}\{l=m\}5 + \mathbbm{1}\{l\neq m\}3,\quad\lambda_{lm}^\ast=\mathbbm{1}\{l=m\}3.0 + \mathbbm{1}\{l\neq m\}1.5.
    \end{split}
\end{align*}
In each scenario, we prepare 50 datasets and conduct repeated experiments.

The strength of overdispersion differs substantially between Scenario~1 and Scenario~2. In fact, in Scenario~1, the mean and variance of \(\mathrm{ZINB}(p_{lm}^\ast, \psi_{lm}^\ast, r_{lm}^\ast)\) are \((40.5, 587.25)\) when \(l=m\) and \((3.6, 48.24)\) when \(l\neq m\). By contrast, in Scenario~2, the mean and variance of \(\mathrm{ZIP}(p_{lm}^\ast,\lambda_{lm}^\ast)\) are \((2.70, 3.510)\) when \(l=m\) and \((0.45, 0.9225)\) when \(l\neq m\).

\subsection{Evaluation metrics}
A deterministic clustering solution \(\hat{z}\) is obtained from the posterior samples of the label vector \(z\). Specifically, we use the variation of information \(\mathrm{VI}\) as the loss function and define \(\hat{z}\) as the label vector minimizing the posterior expected loss \citep{MEILA2007873,WadeGhahramani10.1214/17-BA1073}. Using \(\hat{z}\), we consider the following evaluation metrics: \(\hat{K}\), the number of unique values in \(\hat{z}\); the VI distance \(\mathrm{VI}(\hat{z}, z^{\mathrm{true}})\) between \(\hat{z}\) and \(z^{\mathrm{true}}\); and the VI distance \(\mathrm{VI}(\hat{z}, z_b)\) between \(\hat{z}\) and the boundary of the \(95\%\) credible ball.

\subsection{Results}
Table~\ref{table:sim_results} presents the results of the simulation studies. In both Scenario~1 and Scenario~2, ZINB-SBM accurately recovered \(\true{K}\) and \(\true{z}\). Moreover, the results for \(\mathrm{VI}(\hat{z},z_b)\) indicate that the point estimates obtained by ZINB-SBM are highly accurate. However, in Scenario~1, ZIP-SBM tended to overestimate the number of communities, and \(\hat{z}\) tended to deviate from \(\true{z}\). The larger values and standard deviations of \(\mathrm{VI}(\hat{z},z_b)\) also suggest greater uncertainty, indicating that ZIP-SBM is less stable for highly overdispersed data. In conclusion, ZINB-SBM achieved more robust community detection than ZIP-SBM for overdispersed data.
\begin{table}[tbp]
  \caption{Average values of \(\hat{K}\), \(\mathrm{VI}(\hat{z}, z^{\mathrm{true}})\), and \(\mathrm{VI}(\hat{z}, z_b)\) for ZINB-SBM and ZIP-SBM over 50 Monte Carlo replications under Scenarios~1 and~2. Values in parentheses are standard deviations, and boldface indicates the best result.}
  \label{table:sim_results}
  \centering
    \begin{tabular}{@{}lccccccc@{}}
      \toprule
        & \multicolumn{3}{c}{Scenario~1} && \multicolumn{3}{c}{Scenario~2}\\
        \cmidrule(lr){2-4} \cmidrule(lr){5-8}
        & $\Hat{K}$  & $\text{VI}(\hat{z},\true{z})$ & $\text{VI}(\hat{z},z_b)$ & &$\Hat{K}$  & $\text{VI}(\hat{z},\true{z})$ & $\text{VI}(\hat{z},z_b)$ \\
      \midrule
      ZINB-SBM & $\bm{3} \ (0)$ & $\bm{0}\ (0)$ & $\bm{0}\ (0)$ && $\bm{3}\ (0)$ & $\bm{0}\ (0)$ & $\bm{0}\ (0)$\\
      ZIP-SBM  & $8.620\ (1.640)$ & $1.173\ (0.268)$ & $0.294\ (0.247)$ && $\bm{3}\ (0)$ & $\bm{0}\ (0)$ & $\bm{0}\ (0)$ \\
      \bottomrule
    \end{tabular}
\end{table}

\section{Real data analysis}\label{sec:real}
In this section, we use \texttt{fungusTreeNetwork}, which is available in the \texttt{sbm} package for \texttt{R}. \texttt{fungusTreeNetwork} is a network dataset based on interactions involving 51 tree species and 154 parasitic fungal species. In this paper, we use the tree-to-tree network. Each edge in this network is weighted by the number of parasitic fungal species shared by the two tree species. Genetic distance, taxonomic distance, and geographic distance are available as pairwise covariates for tree-species pairs. See \cite{Vacher_etal2008, mariadassou2010uncovering} for further details. 

We apply CZINB-SBM to the data and compare the results with those of \cite{mariadassou2010uncovering}. Their method is a Bayesian stochastic block model-based on Poisson regression, hereafter referred to as CP-SBM. However, CP-SBM differs substantially from CZINB-SBM in that a regression coefficient is treated as a global parameter rather than block specific ones, and inference is performed by variational methods. Furthermore, using \texttt{fungusTreeNetwork} as a benchmark dataset, we demonstrate that CZINB-SBM achieves higher predictive accuracy for latent interactions than CP-SBM.

\subsection{Results}
Regarding the estimated number of communities \(\hat{K}\), CZINB-SBM yielded \(\hat{K}=3\), while CP-SBM yielded \(\hat{K}=4\). Figure~\ref{fig:Cov} presents the estimated regression coefficients for weighted edges under CZINB-SBM. By contrast, the point estimate of the regression coefficient vector for the pairwise covariates under CP-SBM was \((0.025, -0.564, -0.072)^\top\). 

\begin{figure}[tbp]
  \centering
  \includegraphics[width=\linewidth, height=9.0cm, keepaspectratio]{cov_0422_1.png}
  \caption{Posterior means (\(\circ\) and \(\triangle\)) and 95\% credible intervals of the CZINB-SBM regression coefficients for the pairwise covariates Genetic, Taxonomic, and Geographic distance. Solid lines represent the coefficients \(\beta_{11,1}\), \(\beta_{22,1}\), and \(\beta_{33,1}\) for within-community block pairs, while dashed lines represent the coefficients \(\beta_{12,1}\), \(\beta_{13,1}\), and \(\beta_{23,1}\) for between-community block pairs.}
  \label{fig:Cov}
\end{figure}

The fact that the number of communities estimated by CZINB-SBM is smaller than that supported by CP-SBM may be attributable to the incorporation of block-specific covariate effects in the proposed model. For example, Figure~\ref{fig:Cov} suggests that genetic distance is not, in general, a major factor determining interaction strength overall, whereas it has a strong positive effect specifically within cluster \(2\), that is, for block pair \((2,2)\). Furthermore, taxonomic distance tends to weaken interactions across many block pairs. However, for block pairs \((2,3)\) and \((3,3)\), the uncertainty is large, since the 95\% credible intervals include \(0\). Therefore, we cannot draw a confident conclusion about the direction of the effect of taxonomic distance on edge weights for these pairs. Because CP-SBM is unable to capture such local heterogeneity in covariate effects, it may instead fit the data by subdividing communities. By contrast, the proposed CZINB-SBM provides a useful framework for estimating complex network generating mechanisms in an interpretable manner while accounting for uncertainty via its flexible regression structure and posterior sampling. 

\subsection{Missing link prediction}
Using \texttt{fungusTreeNetwork} as a benchmark dataset, we consider a setting in which some interactions are unobserved (missing), and examine whether the models can accurately predict both the presence and the strength of interactions. To this end, we repeat the following experiment 50 times: 1) Construct the training data \(A^{\mathrm{train}}\) by randomly selecting 20\% of the node pairs \((i,j)\) with nonzero entries \(A_{ij}>0\) and masking them by setting \(A_{ij}=0\). 2) Fit each model using \(A^{\mathrm{train}}\). 3) For all masked pairs \((i,j)\), compute the posterior predictive probabilities \(p(A_{ij} > 0 \mid A^{\mathrm{train}})\) and the posterior predictive expectations \(\E{}{A_{ij}\mid A^{\mathrm{train}}}\). 4) Using the posterior predictive probabilities, compute the area under the curve (AUC), and using the posterior predictive expectations, compute the root mean squared error (RMSE) relative to the true values. 
AUC is a metric for evaluating the performance of binary classification (in this case between \(A_{ij}=0\) and \(A_{ij}>0\)), without depending on the threshold used to classify predictive probabilities into \(0\) or \(1\). Values closer to \(1\) indicate better performance. 

\subsection{Prediction results}
Table~\ref{table:missing} presents the results for AUC and RMSE. According to Table~\ref{table:missing}, CZINB-SBM substantially outperforms CP-SBM in terms of AUC. CZINB-SBM also reduces the RMSE by more than half compared with CP-SBM. These results may be explained by the fact that CZINB-SBM naturally incorporates missing interactions and uses pairwise covariates to complement structural information not directly available from the observations. The improvement in AUC indicates that CZINB-SBM can distinguish missing interactions from non-interactions, while the improvement in RMSE implies that, when an interaction is present, CZINB-SBM can more estimate its weight. Therefore, these results suggest that CZINB-SBM is able to simultaneously recover both the presence and the strength of missing interactions.
\begin{table}[tbp]
  \caption{Average AUC and RMSE values for CZINB-SBM and CP-SBM over 50 Monte Carlo replications. Standard deviations are given in parentheses, and boldface indicates the best result.}
  \label{table:missing}
  \centering
    \begin{tabular}{@{}ccc@{}}
      \toprule
        & AUC & RMSE \\
      \midrule
      CZINB-SBM & $\bm{0.963}\ (0.009)$ & $\bm{2.083}\ (0.428)$ \\
      CP-SBM    & $0.768\ (0.077)$ & $4.525\ (0.556)$ \\
      \bottomrule
    \end{tabular}
\end{table}

\section{Concluding remarks}\label{sec:con}
This study offers two key contributions: proposing ZINB-SBM for flexible community detection in weighted networks under the DMFM framework, and CZINB-SBM for highly accurate interaction prediction and Bayesian inference of regression coefficients with uncertainty quantification. Given the limited research on DMFMs, there is considerable room for both theoretical and applied developments. Finally, developing variational algorithms to scale our framework for large networks represents a promising future direction.

\section*{Acknowledgement}
This work was supported by JST SPRING, Grant Number JPMJSP2132. The authors thank FORTE Science Communications (\url{https://www.forte-science.co.jp/}) for English language editing.
\bibliographystyle{chicago}
\bibliography{cite}

\appendix
\section*{Appendix}
\addcontentsline{toc}{section}{Appendix}
\section{Details of posterior computation for ZINB-SBM}
\subsection{Preliminaries}
In this subsection, we introduce the necessary notation to construct a partially collapsed Gibbs sampler for ZINB-SBM. Let \(X=(x_{ij})\) and \(W=(w_{ij})\) be \(n \times n\) matrices, \(P=(p_{lm})\), \(\Psi=(\psi_{lm})\), and \(R=(r_{lm})\) be \(K \times K\) matrices, and \(z=(z_1,\dots,z_n)\) be the vector of community labels. Here, \(z_{-i}\) denotes the vector \(z\) with \(z_i\) removed, and \(X_{-i}\) and \(W_{-i}\) denote the \((n-1)\times(n-1)\) submatrices of \(X\) and \(W\), respectively, obtained by removing the rows and columns corresponding to node \(i\). Moreover, let \(n_l=\sum_{i=1}^n \mathbbm{1}\{z_i=l\}\), and define
\begin{align*}
x_{lm} &=
\left\{
\begin{array}{ll}
 \sum_{i \neq j; z_i=l,z_j=m}x_{ij}\quad & (l \neq m), \\
 \sum_{i < j; z_i=z_j=l}x_{ij}\quad & (l = m),
\end{array}
\right.
\quad
n_{lm} =
\left\{
\begin{array}{ll}
 n_{l}n_{m}\quad & (l \neq m), \\
 \frac{n_l(n_l-1)}{2}\quad & (l = m),
\end{array}
\right.\\
w_{lm} &=
\left\{
\begin{array}{ll}
 \sum_{i \neq j; z_i=l,z_j=m}w_{ij}\quad & (l \neq m), \\
 \sum_{i < j; z_i=z_j=l}w_{ij}\quad & (l = m).
\end{array}
\right.
\end{align*}
Here, \(x_{lm}\) denotes the total number of connections between communities \(l\) and \(m\), and \(w_{lm}\) denotes the total amount of latent interaction between communities \(l\) and \(m\). The quantities \(x_{lm}^{(-i)}\), \(w_{lm}^{(-i)}\), and \(n_{lm}^{(-i)}\) are defined analogously based on \(z_{-i}\).

In the context of MFMs and DMFMs, care must be taken to distinguish the number of communities \(K\) from the number of occupied clusters \(k\). Here, \(K\) denotes the total number of candidate components, whereas \(k\) denotes the number of components that are actually occupied by at least one node. That is, the model allows empty components. Thus, \(\mathcal{M}_{a}\) denotes the set of occupied community labels, and \(\mathcal{M}_{na}\) denotes the set of empty community labels.

Finally, for \(\mathrm{Gamma}(\gamma,1)\), we define its Laplace transform and the cumulant function associated with the Laplace transform as
\begin{align}\label{eq:LapCum}
    \psi(u) = \left(\frac{1}{1+u}\right)^{\gamma},\quad
    \kappa(u;n) = \gamma^{(n)}\left(\frac{1}{1+u}\right)^{\gamma + n},
\end{align}
where \(x^{(n)}\) denotes the rising factorial, that is, \(x^{(n)}=x(x+1)\cdots(x+n-1).\)
For Bayesian computation of ZINB-SBM, we employ the extended telescoping sampler \citep{iwashige2025bayesianmixturemodelingusing}. Since the extended telescoping sampler is a partially collapsed Gibbs sampler based on an exchangeable partition probability function, the resulting Bayesian computation for ZINB-SBM is also a partially collapsed Gibbs sampler. In addition, the label variables are updated with the block parameters \(P\) and \(\Psi\) marginalized out.

\subsection{Partially collapsed Gibbs sampler for ZINB-SBM}
We summarize the algorithm in Algorithm \ref{alg:ZINB-SBM}. The random variable \(U_n\) in Steps~1 is a latent variable introduced to generalize the mixture weights in DMFMs beyond the Dirichlet distribution, although the Dirichlet distribution is used in this study. It should be noted, however, that the algorithm is formulated in terms of \(S=(S_1,\dots,S_K)\) rather than directly in terms of the mixture weights \(\pi\), where
\[
S_m \mid \gamma, K \sim \mathrm{Gamma}(\gamma/K,1),\quad m=1,\dots,K,
\]
and \(\pi\) is obtained by normalizing \(S\).

In Step~2, we first note the following decomposition:
\begin{align}\label{eq:full_ZINB}
     p(P,\Psi,R,z \mid A,X,W,S,K)
     &=p(P,\Psi,R,z \mid X,W,S,K)\notag\\
     &=p(P\mid X,z,K)p(\Psi\mid R,W,z,K)p(z \mid X,W,R,S,K)p(R \mid W,K).
 \end{align}
Thus, we obtain
\begin{align*}
    p(z \mid X,W,R,S,K) 
    &\propto \int p(X,W,P,\Psi \mid R,S,z,K)\mathrm{d}P\mathrm{d}\Psi \times p(z \mid S,K)\\
    &\propto \int p(X\mid P,z,K)p(P\mid K)p(W \mid \Psi,R,z,K)p(\Psi \mid R,K)\mathrm{d}P\mathrm{d}\Psi \times p(z \mid S,K)\\
    &= p(X\mid z,K)p(W\mid R,z,K)p(z \mid S,K).
\end{align*}
Because of conjugacy with the Beta distribution, \(p(X\mid z)\) and \(p(W \mid R,z)\) can be expressed in closed form using Beta functions. In addition, \eqref{eq:prob_zi_ZINB-SBM} is obtained directly from the following:
\begin{align*}
    p(z_i=c \mid z_{-i}, X, W,S,K) \propto
    S_c \frac{p(X\mid z_i=c,z_{-i},K)p(W\mid R,z_i=c,z_{-i},K)}{p(X_{-i}\mid z_{-i},K)p(W_{-i}\mid R,z_{-i},K)},\quad c=1,\dots,K,
\end{align*}
where $p(X_{-i}\mid z_{-i})$ and $p(W_{-i}\mid z_{-i})$, although constant with respect to $z_i$, are included in the updating equation for $z_i$ for numerical stabilization. 

In Step~2 and~3, block parameters \(P\), \(\Psi\) and \(R\) are updated. From \eqref{eq:full_ZINB}, the full conditional distributions of \(P\) and \(\Psi\) are Beta distributions. To update \(R\), we use a random-walk Metropolis-Hastings step.

In Step~5, the latent variables \(X\) and \(W\) are updated, using the decomposition
\[
p(X,W \mid A,P,\Psi,R,z,S,K)
=
p(W \mid A,X,P,\Psi,R,z)\,p(X \mid A,P,\Psi,R,z).
\]
\eqref{eq:full_w_ZINB-SBM} follows immediately from \(A_{ij}=w_{ij}(1-x_{ij})\). If \(A_{ij}=0\), for \(x_{ij}\), the success probability \(\theta_{ij}\) in \eqref{eq:full_x_ZINB-SBM} is computed as follows:
\begin{align*}
    \theta_{ij} &:= p(x_{ij} = 1 \mid z_i = l, z_j=m, p_{lm}, \psi_{lm}, r_{lm})\\
     &= \frac{p(x_{ij}=1, A_{ij}=0 \mid z_i = l, z_j=m, p_{lm}, \psi_{lm}, r_{lm})}{p(A_{ij}=0 \mid z_i = l, z_j=m, p_{lm}, \psi_{lm}, r_{lm})}\\
     &= \frac{p(x_{ij}=1, A_{ij}=0 \mid z_i = l, z_j=m, p_{lm}, \psi_{lm}, r_{lm})}{p(A_{ij}=0,x_{ij}=0 \mid z_i = l, z_j=m, p_{lm}, \psi_{lm}, r_{lm})+p(A_{ij}=0,x_{ij}=1 \mid z_i = l, z_j=m, p_{lm}, \psi_{lm}, r_{lm})}\\
    &= \frac{p_{lm}}{\psi_{lm}^{r_{lm}}(1-p_{lm})+p_{lm}}.
\end{align*}
If \(A_{ij}>0\), \(x_{ij}=0\).

In Step~6, the hyperparameter \(\gamma\) is updated, where \(p(\gamma)\) denotes the probability density function of \(\mathcal{F}(6,3)\). In Step~7, the number of communities \(K\) is updated according to Proposition~2 of \citet{iwashige2025bayesianmixturemodelingusing}. In Step~8, the unnormalized weights \(S_m\) are updated. If \(K>k\), then Steps~9 and~10 are additionally carried out to update the parameters associated with the empty components. 

It should also be noted that, since the distribution of \(S_m\) depends on \(K\), both the Laplace transform and the cumulant likewise depend on \(K\). In fact, to make this dependence explicit, we write them in the algorithm as \(\psi(u;K)\) and \(\kappa(u;n,K)\), respectively, so that \(\gamma\) in \eqref{eq:LapCum} is replaced by \(\gamma/K\).

\begin{algorithm}[tbp]
\caption{Partially collapsed Gibbs sampler for ZINB-SBM}
\label{alg:ZINB-SBM}
\begin{algorithmic}[htbp]
\STATE\textbf{Step 0.}\ Set initial values.\\
\STATE\textbf{Step 1.}\ Sample $U_{n}$ from $\mathrm{Gamma}(n, T)$,\ where $T = \sum_{m=1}^{M}S_{m}$.
\STATE\textbf{Step 2.}\ Sample $z_{i}$, for $i=1,\dots,n, j=1,\dots,K$ from the following discrete probability distribution defined by each probability
\begin{align}\label{eq:prob_zi_ZINB-SBM}
    p(z_{i} = j\mid\mathrm{rest}) \propto S_{j}\prod_{m=1}^K\frac{B(x_{cm} + a_p, n_{cm} - x_{cm} + b_p)\times B(r_{lm}n_{cm} + a_\psi, w_{cm} + b_\psi)}{B\left(x_{cm}^{(-i)}+a_p, n_{cm}^{(-i)}-x_{cm}^{(-i)}+b_p\right)\times B\left(r_{lm}n_{cm}^{(-i)}+a_\psi, w_{cm}^{(-i)}+b_\psi\right)}.
\end{align}
\STATE\textbf{Step 3.}\ Sample \(p_{lm}\) and \(\psi_{lm}\), for \(1 \leq l \leq m \leq K\), from
\begin{align*}
    p_{lm}\mid \text{rest} \sim \text{Beta}(x_{lm}+a_p,n_{lm}-x_{lm}+b_p),\quad \psi_{lm}\mid \text{rest} \sim \text{Beta}(r_{lm}n_{lm}+a_{\psi},w_{lm}+b_{\psi})
\end{align*}
\STATE\textbf{Step 4.}\ For \(1\leq l \leq m\leq K\), propose \(\log r_{lm}^{(\mathrm{pro})} \sim N(\log r_{lm}, \sigma^2)\), and accept the proposal with probability \(\alpha\), where
\begin{align*}
    \alpha = \min \left\{1, \prod_{i\neq j;z_{i}=l,z_j = m}\frac{\Gamma(w_{ij} + r_{lm}^{(\mathrm{pro})})/\Gamma(r_{lm}^{(\mathrm{pro})})}{\Gamma(w_{ij} + r_{lm})/\Gamma(r_{lm})}\times\frac{\psi_{lm}^{r_{lm}^{(\mathrm{pro})}n_{lm}}p(r_{lm}^{(\mathrm{pro})})r_{lm}^{(\mathrm{pro})}}{\psi_{lm}^{r_{lm}n_{lm}}p(r_{lm})r_{lm}} \right\},
\end{align*}
where, if \(l=m\), the product is taken over \(i<j\) such that \(z_i=z_j=l\).

\STATE\textbf{Step 5.}\ Sample \(x_{ij}\) and \(w_{ij}\), for \(1 \leq i < j \leq n\), from 
\begin{align}
    \begin{split}\label{eq:full_x_ZINB-SBM}
        &x_{ij} \mid \text{rest} \sim \left\{ \,
    \begin{aligned}
    & \delta_0 &\text{if}\ A_{ij} > 0\\
    & \text{Bernoulli}(\theta_{ij}) &\text{if}\ A_{ij}=0\\
    \end{aligned}
\right.\ ,
    \end{split}\\
\begin{split}\label{eq:full_w_ZINB-SBM}
    &w_{ij} \mid \text{rest} \sim \left\{ \,
    \begin{aligned}
    & \delta_{A_{ij}} &\text{if}\ A_{ij} > 0,\\
    & \delta_0 &\text{if}\ A_{ij} = 0, x_{ij}=0\\
    & \text{NB}(\psi_{lm}, r_{lm}) &\text{if}\ A_{ij}=0,x_{ij}=1\\
    \end{aligned}
\right.\ 
\end{split},
\end{align}
where \(\delta_a\) is the Dirac measure at \(a\), and $\theta_{ij} = p_{z_iz_j}/(\psi_{z_iz_j}^{r_{z_iz_j}}(1-p_{z_iz_j})+p_{z_iz_j})$.
\STATE\textbf{Step 6.}\ Propose \(\log \gamma^{(\mathrm{pro})} \sim N(\log \gamma, \sigma^2)\), and accept the proposal with probability \(\alpha\), where
\begin{align*}
    \alpha = \min \left\{1, \left\{\prod_{j=1}^k\kappa(u;n_j,K)\right\}\psi(u;K)^{K-k}p(\gamma) \right\}.
\end{align*}
\STATE\textbf{Step 7.}\ Sample $K$ from the following discrete distribution
\[
\mathbb{P}(K=m\mid\mathrm{rest}) \propto \frac{m!}{(m-k)!}\left\{\prod_{j=1}^{k}\kappa(u;n_j,m)\right\}\psi(u;m)^{m-k}q_K(m),\quad m=k,k+1,\dots.
\]
\STATE\textbf{Step 8.}\ Sample $S_{m}$, for $m \in \mathcal{M}_a$, from \(\text{Gamma}(\gamma/K+n_m, u+1)\)
\STATE\textbf{Step 9.}\ Sample $S_{m}$, for $m \in \mathcal{M}_{na}$, from $ \text{Gamma}(\gamma/K, u+1).$
\STATE\textbf{Step 10.}\ Sample $p_{lm}, \psi_{lm}, r_{lm}$, for $k \leq l \leq m \leq K$ from priors, respectively.
\end{algorithmic}
\end{algorithm}

\section{Details of posterior computation for CZINB-SBM}
\subsection{Preliminaries}
In this subsection, we introduce an overview of P\'{o}lya-Gamma data augmentation. Since 
\begin{align}\label{eq:logistic_ap}
    \psi_{ij} = \frac{\exp(y_{ij}^\top \beta_{z_iz_j,1})}{1+\exp(y_{ij}^\top \beta_{z_iz_j,1})},\quad p_{ij}=\frac{\exp(y_{ij}^\top\beta_{z_iz_j,2})}{1 + \exp(y_{ij}^\top\beta_{z_iz_j,2})},\quad 1 \leq i < j \leq n, 
\end{align}
given $\beta_1, z_i=l,z_j=m, r$, the negative binomial likelihood is as follows:
\begin{align*}
    p(w_{ij}\mid z_i=l,z_j=m,r,\beta_1) 
    &= \frac{\Gamma(w_{ij} + r)}{\Gamma(r)w_{ij}!}(1 - \psi_{ij})^{w_{ij}}\psi_{ij}^r\\
     &= \frac{\Gamma(w_{ij} + r)}{\Gamma(r)w_{ij}!}\frac{\exp(r\eta_{ij,1})}{\left(1 + \exp(\eta_{ij,1})\right)^{w_{ij} + r}},
\end{align*}
where \(\eta_{ij,1} = \text{logit}(\psi_{ij})=y_{ij}^\top \beta_{z_iz_j,1}\). From Theorem~1 in \cite{polson2013bayesian},
\begin{align}\label{eq:PolyGa_w}
    \frac{\exp(r\eta_{ij,1})}{\left(1 + \exp(\eta_{ij,1})\right)^{w_{ij} + r}}
    = 2^{-(w_{ij}+r)}\exp(\kappa_{ij}\eta_{ij,1})\int_{0}^\infty\exp(-\omega_{ij,1}\eta_{ij,1}^2/2)p(\omega_{ij,1}\mid w_{ij}+r,0)\mathrm{d}\omega_{ij},
\end{align}
where $\kappa_{ij,1} = (r-w_{ij})/2$ for \(i \neq j\), and we set $\kappa_{ii,1} = 0$ for \(i = j\), since self-loops are not considered. Here, $p(\omega\mid b, 0)$ is a density of P\'{o}lya-Gamma distribution $\text{PG}(b, 0)$. In addition, the integrand \(\exp(-\omega_{ij,1}\eta_{ij,1}^2/2)\,p(\omega_{ij,1}\mid w_{ij}+r,0)\)
is proportional to the density of \(\mathrm{PG}(w_{ij}+r,\eta_{ij,1})\), since it is obtained by exponential tilting of \(\mathrm{PG}(w_{ij}+r,0)\). Here, we set \(\omega_{ii,1}=0\). In the same way, given $\beta_2, z_i=l,z_j=m, r$, we have
\begin{align*}
   p(x_{ij}\mid z_i=l,z_j=m,\beta_2) 
   &= p_{ij}^{x_{ij}}(1-p_{ij})^{1-x_{ij}}\\ 
   &= \frac{\exp(\eta_{ij,2})^{x_{ij}}}{1 + \exp(\eta_{ij,2})}
\end{align*}
where \(\eta_{ij,2} = \text{logit}(p_{ij})=y_{ij}^\top \beta_{z_iz_j,2}\). From Theorem~1 in \cite{polson2013bayesian},
\begin{align}\label{eq:PolyGa_x}
    \frac{\exp(\eta_{ij,2})^{x_{ij}}}{1 + \exp(\eta_{ij,2})} = 2^{-1}\exp(\kappa_{ij,2}\eta_{ij,2})\int_{0}^{\infty}\exp(-\omega_{ij,2}\eta_{ij,2}^2/2)p(\omega_{ij,2}\mid1,0)\mathrm{d}\omega_{ij,2},
\end{align}
where where $\kappa_{ij,2} = x_{ij}-1/2$ for \(i \neq j\), and we set $\kappa_{ii,2} = 0$ for \(i = j\). In addition, the integrand \(\exp(-\omega_{ij,2}\eta_{ij,2}^2/2)\,p(\omega_{ij,2}\mid 1,0)\)
is proportional to the density of \(\mathrm{PG}(1,\eta_{ij,2})\), since it is obtained by exponential tilting of \(\mathrm{PG}(1,0)\). Here, we set \(\omega_{ii,2}=0\).

Finally, for pairs \((i,j)\) satisfying \(z_i=l\) and \(z_j=m\), we introduce the following notations. Here, \(s=1\) refers to quantities associated with \(w_{ij}\), whereas \(s=2\) refers to quantities associated with \(x_{ij}\). Let \(\beta_s \in \mathbb{R}^{K \times K \times q}\) denote the tensor of regression coefficients indexed by block pairs, and \(\Omega_s \in \mathbb{R}^{n \times n}\) denote the symmetric matrix of P\'{o}lya-Gamma latent variables. If \(n_{lm} \geq 1\), let \(Y_{lm}\in \mathbb{R}^{n_{lm}\times q}\) denote the \(lm\) block design matrix whose rows are given by \(y_{ij}^{\top}\), let \(\kappa_{lm,s}\in \mathbb{R}^{n_{lm}}\) denote the vector whose entries are \(\kappa_{ij,s}\), and let \(\Omega_{lm,s}\in \mathbb{R}^{n_{lm}\times n_{lm}}\) denote the diagonal matrix whose diagonal entries are \(\omega_{ij,s}\). If \(n_{ll}=0\), that is, if community \(l\) consists of a single \(i\)th node, then \(Y_{ll}\) is the \(q\)-dimensional zero vector and \(\Omega_{ll,s}=\omega_{ii,s}=0\) is the zero matrix.

\subsection{Partially collapsed Gibbs sampler for CZINB-SBM}
Algorithm~\ref{alg:CZINB-SBM} is a partially collapsed Gibbs sampler for CZINB-SBM. Here, Steps~1,~6,~7,~8, and~9 are omitted, since they are the same as those in Algorithm~\ref{alg:ZINB-SBM}. In Steps~2 and~3, we note the following decomposition:
\begin{align*}
    p(\beta_1,\beta_2,z \mid A,X,W,r,\Omega_1,\Omega_2,S,K)
    &=p(\beta_1,\beta_2,r,z \mid X,W,r,\Omega_1,\Omega_2,S,K)\notag\\
    \begin{split}
    &=p(\beta_1 \mid W, z, r, \Omega_1,K)p(\beta_2 \mid X, z, \Omega_2,K)\times\\
    &p(z \mid X, W, r,\Omega_1, \Omega_2, S, K)
    \end{split}.
\end{align*}
Thus, we obtain
\begin{align*}
    p(z \mid X, W, r,\Omega_1, \Omega_2, S, K)
    &\propto \int p(X,W,\beta_1,\beta_2 \mid r,\Omega_1, \Omega_2,S,z,K)\mathrm{d}\beta_1\mathrm{d}\beta_2 \times p(z \mid S,K)\\
    &\propto \int p(W\mid \beta_1,r,\Omega_1,z,K)p(\beta_1\mid K)p(X\mid \beta_2,\Omega_2,z,K)p(\beta_2 \mid K)\mathrm{d}\beta_1\mathrm{d}\beta_2 \times \\
    &p(z \mid S,K)\\
    &= p(W\mid r,\Omega_1,z,K)p(X\mid \Omega_2,z,K)p(z \mid S,K).
\end{align*}
Because full conditional distributions of \(\beta_s\) are normal distributions, \(p(W\mid r,\Omega_1,z,K)\) and \(p(X\mid \Omega_2,z,K)\) can be expressed in closed form. In fact, the full conditional distribution of $\beta_{lm,s}$ is
\begin{align*}
    p(\beta_{lm,s} \mid \text{rest}) 
    &\propto\exp\left(-\frac{1}{2}\beta_{lm,s}^\top Y_{lm}^\top\Omega_{lm,s}Y_{lm}\beta_{lm,s} + \beta_{lm,s}^\top Y_{lm}^\top\kappa_{lm,s}\right)p(\beta_{lm,s}).
\end{align*}
Since \(\beta_{lm,s}\sim \mathcal{N}(b_0,B_0)\) and, for \(c=1,\dots,K\),
\[
p(z_i=c \mid z_{-i}, X, W,r,\Omega_1,\Omega_2,S,K) \propto
    S_c \frac{p(W\mid r,\Omega_1,z_i=c,z_{-i},K)p(X\mid \Omega_2,z_i=c,z_{-i},K)}{p(W_{-i}\mid r,\Omega_{-i,1}z_{-i},K)p(X_{-i}\mid \Omega_{-i,2},z_{-i},K)},
\]
we obtain easily \eqref{eq:prob_zi} and \eqref{eq:full_beta_s}.

In Step~4, ~5a, and~5b, the following decomposition is used:
\[
p(X,W,r \mid A,\beta_1,\beta_2,z,S,K)
=
p(W \mid A,X,\beta_1,r,z)\,p(X \mid A,\beta_1,\beta_2,r,z)p(r).
\]
In Step~4, we use a random-walk Metropolis-Hastings step to update the global parameter \(r\), and, in Step~5a and~5b, the latent variables \(X\) and \(W\) are updated. The derivation in \eqref{eq:full_w_CZINB-SBM} and \eqref{eq:full_x_CZINB-SBM} is essentially the same as that in \eqref{eq:full_w_ZINB-SBM} and \eqref{eq:full_x_ZINB-SBM}. However, note that Algorithm~\ref{alg:ZINB-SBM} is written in terms of \(\psi_{lm}, p_{lm}, r_{lm}\), whereas Algorithm~\ref{alg:CZINB-SBM} is written in terms of \(\psi_{ij},p_{ij}, r\). The update of \(\omega_{ij,s}\) in Step~5b follows directly from \eqref{eq:PolyGa_w} and \eqref{eq:PolyGa_x}. In Step~10, the regression coefficients for the empty components are sampled from the prior. 

\begin{algorithm}[tbp]
\caption{Partially collapsed Gibbs sampler for CZINB-SBM}
\label{alg:CZINB-SBM}
\begin{algorithmic}[htbp]
\STATE\textbf{Step 2.}\ Sample $z_{i}$, for $i=1,\dots,n, j=1,\dots,K$ from the following discrete probability distribution defined by each probability
\begin{align}\label{eq:prob_zi}
    p(z_{i} = j\mid\mathrm{rest}) \propto S_{j}&\prod_{s=1}^2\prod_{m=1}^{K}\frac{|B_{cm,s}|^{1/2}\exp\left(\frac{1}{2}b_{cm,s}^\top B_{cm,s}^{-1}b_{cm,s}\right)}{|B_{cm,-i,s}|^{1/2}\exp\left(\frac{1}{2}b^{^\top}_{cm,-i,s} B_{cm,-i,s}^{-1}b_{cm,-i,s}\right)},
\end{align}
where
\[
B_{lm,s} = (B_0^{-1} + Y_{lm}^\top \Omega_{lm,s}Y_{lm})^{-1}, b_{lm,s} = B_{lm,s}(B_0^{-1}b_0 + Y_{lm}^\top \kappa_{lm,s}),\quad s=1,2.
\]
\STATE\textbf{Step 3.}\ Sample \(\beta_{lm,s}\), for \(1 \leq l \leq m \leq K\) and \(s=1,2\), from
\begin{align}\label{eq:full_beta_s}
    &\beta_{lm,s} \mid \mathrm{rest} \sim \mathcal{N}_q(b_{lm,s}, B_{lm,s}),
\end{align}
\STATE\textbf{Step 4.}\ Propose \(\log r^{(\mathrm{pro})} \sim N(\log r, \sigma^2)\), and accept the proposal with probability \(\alpha\), where
\begin{align*}
    \min \left\{1, \prod_{i < j}\frac{\Gamma(w_{ij} + r^{(\text{pro})})/\Gamma(r^{(\text{pro})})}{\Gamma(w_{ij} + r)/\Gamma(r)}\psi_{ij}^{r^{\text{(pro)}} - r}\times \frac{p(r^{\text{(pro)}})r^{\text{(pro)}}}{p(r)r} \right\}.
\end{align*}

\STATE\textbf{Step 5a.}\ Sample \(x_{ij}\) and \(w_{ij}\), for \(1 \leq i < j \leq n\), from
\begin{align}
\begin{split}\label{eq:full_x_CZINB-SBM}
    &x_{ij} \mid \text{rest} \sim \left\{ \,
    \begin{aligned}
    & \delta_0 &\text{if}\ A_{ij} > 0\\
    & \text{Bernoulli}(\theta_{ij}) &\text{if}\ A_{ij}=0\\
    \end{aligned}
\right.\ 
\end{split},\\
\begin{split}\label{eq:full_w_CZINB-SBM}
    &w_{ij} \mid \text{rest} \sim \left\{ \,
    \begin{aligned}
    & \delta_{A_{ij}} &\text{if}\ A_{ij} > 0,\\
    & \delta_0 &\text{if}\ A_{ij} = 0, x_{ij}=0\\
    & \text{NB}(\psi_{ij}, r) &\text{if}\ A_{ij}=0,x_{ij}=1\\
    \end{aligned}
\right.\
\end{split},
\end{align}
where \(\delta_a\) is the Dirac measure at \(a\), and $\theta_{ij} = p_{ij}/(\psi_{ij}^{r}(1-p_{ij})+p_{ij})$.
\STATE\textbf{Step 5b.}
Sample \(\omega_{ij,1}\) and \(\omega_{ij,2}\), for \(1 \leq i < j \leq n\), from
\begin{align*}
    \omega_{ij,1} \mid \text{rest} \sim \text{PG}(w_{ij} + r, y_{ij}^\top\beta_{lm,1}),\quad
    \omega_{ij,2} \mid \text{rest} \sim \text{PG}(1, y_{ij}^\top\beta_{lm,2}).
\end{align*}
\STATE\textbf{Step 10.}\ Sample $\beta_{lm,1}, \beta_{lm,2}$, for $k \leq l \leq m \leq K$ from priors, respectively.
\end{algorithmic}
\end{algorithm}

\end{document}